\documentclass[conference]{IEEEtran}
\makeatletter
\def\ps@headings{%
\def\@oddhead{\mbox{}\scriptsize\rightmark \hfil \thepage}%
\def\@evenhead{\scriptsize\thepage \hfil \leftmark\mbox{}}%
\def\@oddfoot{}%
\def\@evenfoot{}}
\makeatother
\usepackage{subfigure}
\usepackage{latexsym}
\usepackage{graphicx}
\usepackage{verbatim}

\begin{document}

\title{Defeating Internet attacks and Spam using ``disposable'' Mobile IPv6 home addresses}
\author{\large Pars Mutaf\\
    \normalsize Ege University International Computer Institute\\
    \normalsize Izmir, Turkey\\
    \normalsize pars.mutaf@ege.edu.tr\\}
\date{}
\maketitle

\begin{abstract}
We propose a model of operation for next generation wireless Internet, 
in which a mobile host has hundreds of ``disposable'' Mobile IPv6
home addresses. Each correspondent is distributed a different
disposable home address. If attacked on a given home address, the mobile
user can block packets to that address and become unreachable to the 
attacker. Blocking one address does not affect other addresses. Other
correspondents can still reach the mobile host. A new home address
can also be requested via e-mail, instant messaging, or directly 
from the target host using a protocol that we develop. This model is 
especially useful against battery exhausting Denial-of-Service (DoS) 
attacks and CPU exhausting distributed DoS attacks, since it seems to 
be the only viable solution, currently. We show however that this model 
can also be used to defeat other attacks and also to stop spam. 
\end{abstract}

\section{Introduction}

In the next generation Internet, it is estimated that billions
of hand-held IP-addressable mobile phones will be directly connected
to the wireless edges of the Internet, obseleting today's cellular 
telephony architectures. These devices will be equipped with VoIP 
(Voice over IP) software and standard IP applications e.g. Web 
browsing, instant messaging, file transfer etc. Users will enjoy 
constant wireless Internet connection regardless of their location 
i.e., not only in hotspots, but also in urban and rural areas.

In order to support such a large number of Internet hosts (mostly
wireless and mobile) the next generation internet protocol, i.e., 
the IETF (Internet Engineering Task Force) standard IPv6 which 
provides 128-bit addresses will probably be used \cite{IPv6}. 
A mobility management solution will also be needed to help mobile 
hosts receive packets regardless of their location i.e. current 
subnet. The IETF standard Mobile IPv6 (MIPv6) is currently viewed as 
the strongest candidate to address the mobility management problem\cite{MIPv6}. 

In MIPv6, the mobility is handled at the IP layer. Mobile hosts are
IP-addressable and are constantly connected to the Internet.
The Internet is, however, a hostile environment where malicious parties
can easily eavesdrop voice and data traffic, impersonate other users, 
mount Denial-of-Service (DoS) attacks, infect them with viruses and
worms and mount other attacks. Encryption and authentication solutions 
are available to defeat eavesdroppers and impersonators, however DoS 
attacks are particularly difficult to cope with since the Internet was 
designed to route any packet to its destination. Malicious packets
can be sent to a destination in order to consume its resources, e.g.
energy and CPU cycles. These attacks that we call 
\textit{battery/CPU exhaustion} 
attacks will be particularly important as far as mobile 
users are concerned, since mobile hosts are battery-powered and their 
CPU power is limited. Attackers may also exploit buggy code to obtain 
unauthorized access or degrade performance, or infect victim hosts 
with viruses and worms, etc. Finally, SPam over IP Telephony (SPIT) 
will also be a potential problem in next generation Internet.

In this paper, we propose a solution to these attacks. Our proposal is 
called \textit{disposable home addresses} (see next section for a 
review of Mobile IPv6) and inspired from the disposable phone numbers 
and e-mail addresses that are in use today. In our proposal, MIPv6 
home addresses are used as phone numbers. A mobile user distributes 
a different disposable home address to each of their correspondents. 
If attacked on one of the home addresses, the compromized address is 
canceled, i.e. disposed of. Other correspondents that were distributed 
other home addresses are not affected. They can still reach the mobile 
host. New disposable home addresses can be remotely requested by e-mail 
or instant messaging, however, we also develop a protocol for 
directly requesting a disposable home addresses form the target mobile 
host. This model is especially useful against battery exhausting 
DoS attacks and CPU exhausting distributed DoS attacks, since it seems 
to be the only viable solution, currently. We show however that this 
model can also be used to defeat other attacks and also to stop spam. 

The rest of this paper is organized as follows: Section \ref{sec:ps}
describes the problem that we address in this paper, Section 
\ref{sec:defeat} describes the disposable home address solution 
that we propose, Section \ref{sec:dist} proposes a protocol for distributing
home addresses, Section \ref{sec:certificateless} proposes a solution
to the same problem without assuming the existence of a Public Key
Infrastructure (PKI), Section \ref{sec:spit} proposes a solution
to the SPIT (SPam over IP Telephony) problem, 
Sections \ref{sec:arch}, \ref{sec:lim}, \ref{sec:imp}
provide discussions of some architectural considerations, some limitations
of our proposal and implementation considerations and finally Sections
\ref{sec:related} and \ref{sec:conc} present related work and conclusions.

\section{Mobile IPv6}

\begin{figure}[t]
	\centering
 	\includegraphics[width=7cm]{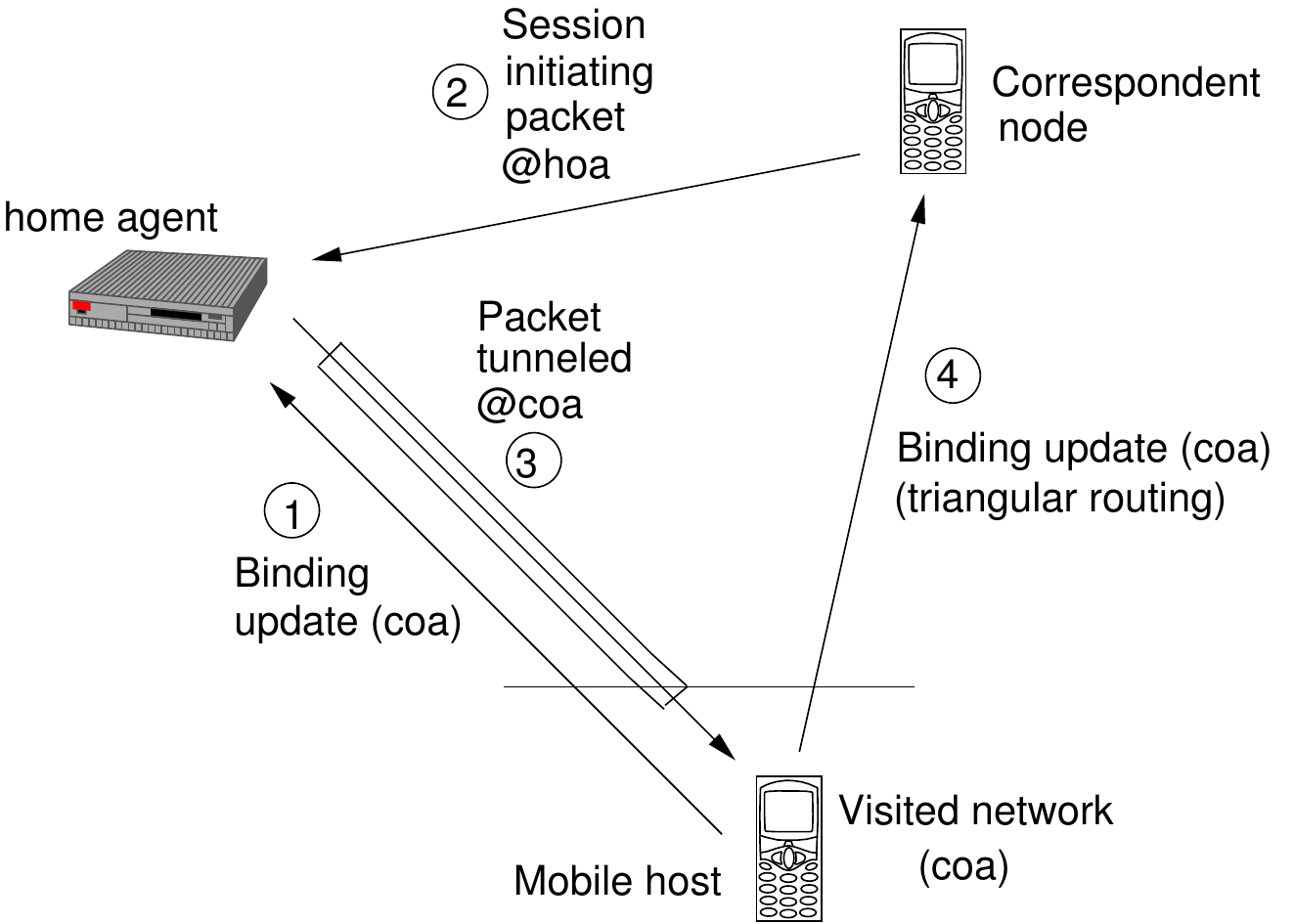}
	\caption{MIPv6 model of operation (simplified).}
	\label{fig:mipv6}
\end{figure}

The Mobile IPv6 (MIPv6) model of operation is illustrated in 
Figure \ref{fig:mipv6}. 
In MIPv6, a mobile host is assigned two IP addresses. 
The first one is called a \textit{home address} (hoa). It is a fixed address 
that points to the mobile host's \textit{home network}, where a 
special router called \textit{home agent} serves the mobile host. 
When away from home, each time a mobile host visits a new subnet, it 
configures a new address called a \textit{care-of-address} (coa). 
The mobile host then sends a \textit{binding update} message to its home agent 
reporting its current care-of-address. Consequently, the home agent 
always knows the current IP address of the mobile host. 

When a correspondent node needs to establish a session with the mobile
host, it sends a packet to the mobile host's fixed address which is the
home address. This packet is intercepted by the home agent and tunneled
to the mobile host's current care-of-address. Upon receipt of the 
packet, the mobile host decapsulates the packet and obtains the original 
packet sent by the correspondent node. The mobile host has now the
IP address of the correspondent node and sends a binding update message
to the correspondent node. The correspondent node consequently learns the
current IP address of the mobile host and the two hosts
can communicate directly without the help of the home agent. This 
optimization is called \textit{route optimization} or \textit{triangular
routing}. An alternative mode of operation is called \textit{bidirectional tunneling},
in which a binding update is not send to the correspondent node. The 
mobile hosts's reply is reverse tunneled back to the home agent which decapsulates
the packet and sends it to correspondent node. This packet's source
address is the mobile host's home address. This mode of operation is 
generally prefered for location privacy. The location i.e. current 
care-of-address of the mobile host is hidden from the correspondent node.

\section{Problem statement}
\label{sec:ps}

In this section, we present several Internet attacks that can be mounted
against next generation mobile hosts. We especially focus our attention
on battery/CPU exhaustion attacks since currently disposable home addresses 
seem to be the only defense against these attacks. However, other
types of Internet attacks e.g. port scanning, attacks that exploit buggy 
code, spam or any future attack that we cannot foresee can also be defeated 
using the proposed solution. 

By current practice, it is assumed that a higher-layer identifier e.g. a
Fully Qualified Domain Name (FQDN)\cite{DNS} or if (Session 
Initiation Protocol)\cite{SIP} is used, a SIP URI (Uniform Resource Locator)
will be resolved to a home address and the destination mobile host will be 
reachable at that address. Consequently, an attacker who learned these
identifiers can mount attacks against the obtained home address.

One solution to these threats is to keep these higher-layer 
identifiers secret and not sending them in clear-text over the
Internet (the latter being a pre-caution against eavesdroppers). We 
feel however uncomfortable with this approach. Mobile host identifiers 
need to be shared for communication and we believe that they may 
easily leak to the hands of malicious users and attackers.

\subsection{Battery exhaustion attacks}

In wireless communications, energy is a scarce resource since terminals
are battery powered. Most link-layer wireless access protocols support 
a mode called \textit{dormant mode}, or \textit{power save mode} in 
which the network interface consumes much less energy. In this mode, 
the receiver circuits are mostly off, and they are periodically switched on to 
check if a packet was received while the terminal was in dormant mode, 
and buffered by the access point or base station. This optimization
is very important because listening consumes an important amount of
energy and terminals are idle (i.e. not sending or receiving) most of
the time. Transmitting consumes even more energy, and in power
save mode, terminals do not transmit frames.

In cellular access technologies, dormant mode is possible in a much 
larger area than a single cell. Cells are grouped into paging areas,
i.e., each \textit{paging area} contains a large number of cells and covers
a large geographical area. When moving within paging area boundaries, 
a terminal remains in dormant mode and does not report its exact 
location (cell) to the network and conserves energy. When there is
an incoming call, however, the network needs to page the terminal
in its current paging area. This is achieved by broadcasting a 
\textit{paging message} in all cells of the paging area. The terminal 
then wakes up, reports its exact location and the call is delivered.  

An attacker (located anywhere in the Internet) who learned the home
address of a mobile host, can mount a DoS attack by continuously 
sending packets to that address. This attack would consume the energy 
of the target mobile host by preventing it from entering power save mode 
and forcing it to send reply packets (and possibly link-layer ACKs 
returned to the leaf router or access point). This is a fundamentally 
difficult problem. It needs addressing because an important source of 
energy consumption in a battery powered mobile device is the network 
interface. Note that the energy consumption problem here is two-fold:

\begin{itemize}
\item Victim is deprived from sleep mode (network interface)
\item Victim is forced to transmit reply packets (link-layer ACKs and 
upper layer reply packets e.g. TCP SYN/ACK, or SIP OK)
\end{itemize}

It can be noted that two kinds of attacks are possible. In the first
one, the attacker may only want to deprive the victim from power save
mode. Let \(T\) the time duration a victim waits before entering sleep 
mode when it is not in active communication. The victim can be deprived from 
sleep mode by sending \(1/T\) packets per second. In the second attack,
the attacker can send more frequent packets and force the victim to 
send reply packets and consume more energy in addition to sleep mode 
deprivation. In this paper, we assume the second type of attack. The
attacker sends as much packets as possible in order to exhaust the
victim's battery. If the attack succeeds, the device may become 
completely unavailable until its battery is recharged. The user who 
may not have foreseen such premature battery outage may not have their 
charger with them, or a place to recharge the battery may not be 
immediately available. 

Experiments showed that an infrastructure mode 802.11 equipped 
HTC mobile phone's energy (full battery, 1350 mAh) can be consumed in \(\sim 3.5-4\) 
hours by continuously sending it ICMP echo request (ping) 
packets\footnote{In these experiments the \textit{Linux} command \textit{ping -f target} 
was used, which sends a new ping packet as soon as a reply is received 
to the previous one, i.e. without waiting one second. In practice the mobile
phone may silently drop ping packets. However, an attacker can also send
other request packets e.g. TCP SYN, or SIP INVITE which cannot be silently
dropped, otherwise the target may deny service to legitimate session 
initiators. For experimental purposes sending ICMP echo requests is not
different from sending TCP SYN, or SIP INVITE. The impact on the target
phone is the same. Using the ping command also allows us to test the effect 
of different payload sizes.}. Under normal conditions, i.e. without attack, 
the phone survives \(\sim 1 \) day without recharging (802.11 activated 
but idle). Note also that the attacked phone's battery was full. The attack 
may shut down a target mobile device much more quickly if its battery 
level is low. Attackers can probably obtain better results against 
outdoor technologies. Cellular terminals usually have larger distances 
to the base stations, requiring more power to overcome low signal-to-noise 
ratio\cite{ENCONS}. I.e., replies to malicious packet will need to be
transmitted at a higher power, consuming the victim's battery more
quickly.

Note that a simple solution e.g. filtering the attacker's IP 
address would not work because the attacker can randomly spoof source 
IP addresses. End-to-end authentication solutions (e.g. IPsec\cite{IPsec}) also 
do not solve this problem, since the packet is dropped by the target, 
which is too late. Energy is consumed for receiving the packet,
possibly generating a link-layer ACK, and the target is deprived from
link-layer power save mode since it is continuously receiving (and 
dropping) packets.

\subsection{CPU exhaustion attacks}

CPU exhaustion attacks aim to consume a victim host's CPU cycles by
continuously sending it requests e.g. ICMP echo request, TCP SYN, SIP 
INVITE etc. The victim host will continuously process the requests, 
prepare replies and send them back to fake sender address(es). 
Depending on its CPU power, the victim's performance may be degraded 
to the point that it cannot be used during the attack. 

Note that if IPsec is used, packets from untrusted hosts will be 
quickly dropped without consuming more CPU cycles\cite{IPsec}. However, 
this assumes that the target host has an IPsec security association 
with every trusted host in the Internet. 

More importantly, an attacker can exploit the Internet Key Exchange 
(IKEv2) protocol of the IPsec protocol suite\cite{IKEv2}. 
In IKEv2, CPU expensive cryptographic computations are made in response 
to an IKEv2 initiation request. A well known DoS attack consists of sending 
an outstanding number of IKEv2 requests and consume the victim's CPU 
cycles. IKEv2 supports cookies in order prevent an attacker from 
spoofing their IP address. In response to an IKEv2 initiation request, 
the responder returns a cookie in order to check the return routability
of the initiator. Only if the initiator used its real IP address
it will receive the cookie. A legitimate initiator then repeats its 
request with the received cookie. The responder, concluding that the
initiator is legitimate (i.e. not spoofing IP addresses), proceeds 
with CPU expensive cryptographic
computations that are necessary for key agreement. Consequently, an 
attacker is forced to use their real network address which can be 
blocked if frequent requests are made from the same address. 

Note however that an attacker can organize a Distributed DoS (DDoS)
attack in which thousands of compromized hosts attack using their real 
address (the attack would probably target many mobile hosts).
In this case, cookies would be useless since attacking hosts use their
real address and blocking thousands of addresses without denying 
service to legitimate initiators is a difficult task. A different 
solution is therefore necessary.

\subsection{Other attacks}

Other attacks may also be mounted against future mobile hosts. For 
example attackers may port scan a target host to find
about vulnerabilities, try to obtain unauthorized access or infect the 
target host with a virus or worm. 

\subsection{SPIT: SPam over IP Telephony}

Similarly to e-mail, VoIP systems are likely to be abused by malicious
parties who initiate unsolicited and unwanted communications. 
Telemarketers, prank callers, and other telephone system abusers are 
likely to target VoIP systems. This problem is refered to as SPam over 
IP Telephony (SPIT)\cite{SPIT}.

Unlike battery and CPU exhaustion attacks, this threat is an 
application-layer threat and disturbs the user in a different way. We 
show in this paper that the same solution that we develop against 
battery/CPU exhaustion attacks can be used to fight SPIT. 

\subsection{Location privacy attacks}

An attacker located anywhere in the Internet can periodically send packets to 
a victim's home address without establishing any session, forcing them 
to send unnecessary binding updates (in route optimization mode) and 
reveal their location. 

\section{Disposable home addresses}
\label{sec:defeat}

In the proposed model, a mobile host generates a large number 
of random \textit{disposable home addresses} pointing to its same home 
network by sending a \textit{home address request} message to its home 
agent\footnote{In MIPv6
a mobile host has an IPsec security association with its home agent, hence
home address generation is secure. When generating home addresses, 
address collisions can also be detected by the home agent which has a 
list of all home addresses in the home network.}. 
Each time Alice is requested her identifier (a home address 
which is in this case equivalent to a phone number) by another user, 
she picks a different home address generated by her device and tells 
it to the peer user if both users are present (refered to as ``user contact'') 
or sends via e-mail if the request was made by e-mail. 

When the user moves from one network to another, it configures a new 
care-of-address as defined in \cite{MIPv6} and sends a binding update for 
one of its home addresses. In our case, a binding update sent for one home 
address must update the care-of-address binding for all of the home 
addresses of the mobile host. I.e., the mobile host can receive an 
incoming session to any home address that it previously configured and 
distributed.

Let \(hoa_i\) be one of the home addresses of Alice's battery-powered
mobile device. When idle, i.e. not in active communication, Alice's 
device is assumed to enter \textit{power save mode} in which the network 
interface consumes much less power. As described above an attacker who
learned the address \(hoa_i\) may attack that address. The attacker may 
remotely consume the victim's energy and/or CPU cycles or mount other 
attacks as described above.

Since Alice has distributed a different home address to each host
in her address book, she can block the packets addressed to \(hoa_i\).
Upon detecting the attack (receipt of an unusually large number of 
packets to \(hoa_i\)), Alice's device needs to send a message to its 
home agent to indicate that packets destined to \(hoa_i\) should be 
dropped (upon user command). 
Alice's device should also configure a new care-of-address 
(and send to its home agent in a binding update message) 
if the previous one was sent to the 
attacker for triangular routing. Clearly, the attacker's packets will 
not reach Alice's device anymore. If in active communication with a
legitimate correspondent, that node must also be sent the new
care-of-address using a binding update message. 

If Alice receives an incoming session from another user, the session
will be destined to the home address \(hoa_j\), which will be allowed. Blocking one home  
address does not affect the incoming sessions to other home addresses.
Note also that the attacker has no knowledge of the home addresses 
distributed to other users, thus cannot easily attack another 
address. When the attack on \(hoa_i\) stops, Alice can reactivate that 
address. Alternatively, the correspondent user that were provided 
with address \(hoa_i\), can request another address by e-mail, instant messaging or using
the home address distribution protocol that we describe in Section \ref{sec:dist}.

An IPv6 address is 128-bit long: 64 bits network address and 64 bits 
interface identifier.  The total number of home addresses in a home 
network cannot exceed \(2^{64}\). This is not an important problem 
since \(2^{64}\) is a huge number. For example, if 10,000 mobile 
hosts are served by the same home network and each host has 200 peers 
in their address book, hence 200 home addresses, 
then \(  \frac{2 \times 10^6}{2^{64}}   \simeq 2 \times 10^{-12} \) of the 
available address space in the home network will be in use.  

\section{Distributing disposable home addresses}
\label{sec:dist}

\subsection{Protocol description}

As described above, in our proposal, home addresses are used as phone
numbers (disposable ones).
A disposable home address can be shared when there is user contact by 
verbal communication, or remotely by e-mail, instant messaging etc. In 
this section, we develop a protocol for requesting a disposable home 
address from a target host directly. This protocol will run between 
peer contact manager applications (see below for benefits). 

In this model, each mobile host has a \textit{prime home address}. A
disposable home address is first requested from the target's prime 
home address and the call is made to the returned disposable home 
address. Incoming calls to the prime home address are not accepted, 
and an error message is returned by the target host indicating that a 
disposable home address must be requested. The prime home address is 
registered to the DNS and mobile hosts (and hence their users) are
identified by the corresponding FQDN (Fully Qualified Domain Name). 
Users can share their FQDN when there is user contact through verbal 
communication or remotely by e-mail, instant messaging etc. 

\begin{figure}[t]
	\centering
 	\includegraphics[width=7cm]{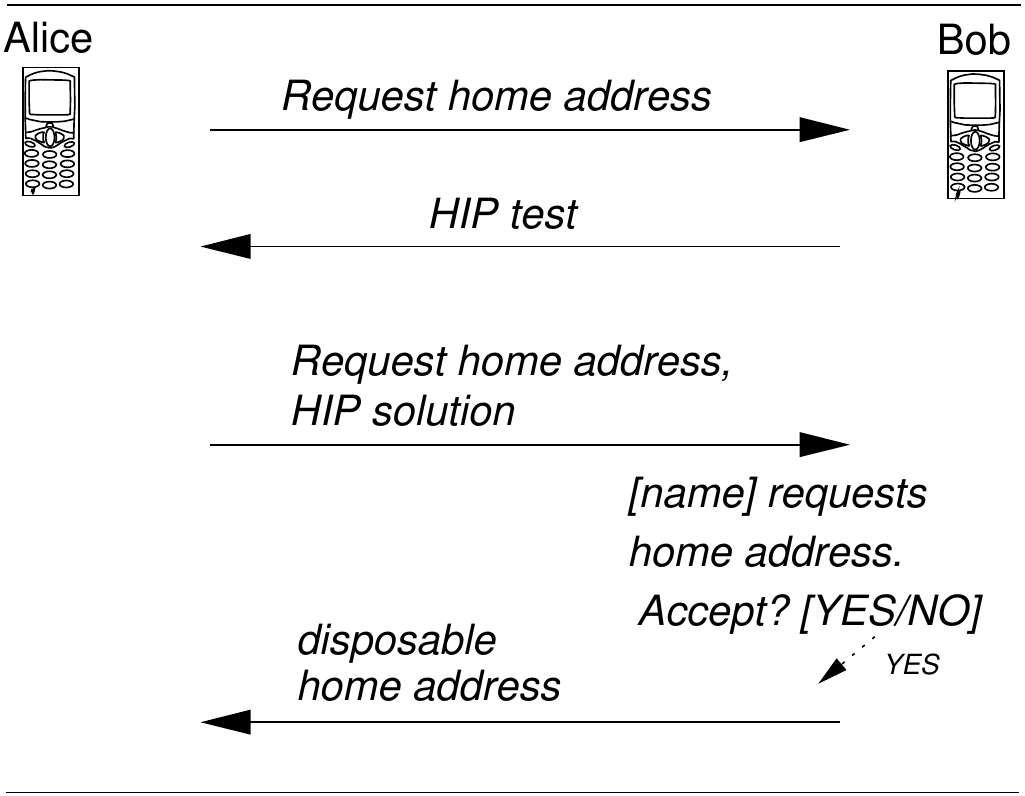}
	\caption{Protocol description (details omitted).}
	\label{fig:protocol}
\end{figure}

Using the proposed protocol, illustrated in Figure \ref{fig:protocol}, 
Alice's phone can remotely request a disposable home address from 
Bob's phone which is located anywhere in the Internet. The same 
protocol can be used when there is user contact, i.e. Alice and Bob 
are both present. Alice enters Bob's 
FQDN to her contact manager application which makes a disposable home 
address request to the target mobile phone's prime home address. The request contains 
Alice's name and possibly other information about Alice, and if 
possible the request should be signed with Alice's private key and her 
certificate must be attached. Upon receipt of the request, the peer 
application on the target phone displays a message indicating that 
Alice requests a disposable home address and ask Bob's permission to 
return the requested address. If Bob accepts, the address is returned 
and Alice's phone can make a call to that address. Bob's reply containing the 
disposable home address must be signed using Bob's private key, and 
Bob's certificate must be attached. Note that Alice does not deal 
with disposable home addresses. She simply makes a call to Bob in her 
address book and the contact manager requests a disposable home 
address if necessary and makes the call. The call is made to the 
disposable home address. If the disposable home address given to Alice 
is compromized and used for an attack, Bob will block this address as 
previously described. In this case, Alice's phone will request a new 
home address when needed. 

This protocol brings the following advantages:

\begin{enumerate}

\item Ease of use: Users are less concerned with disposable home 
addresses since they use FQDNs. They may even ignore that they need 
one to reach the target host (see below for exceptions). User enters
the target FQDN to the contact manager which requests a disposable
home address from the target phone if needed, and makes a call to that address.

\item Speed and reliability: Using the protocol described in this 
section, a disposable home address can be requested quickly and more 
reliably. Using e-mail, the requesting user needs to wait until the 
target user checks e-mail (if push e-mail is not used) and the e-mail 
servers may be temporarily down, unreachable or congested. Using the 
proposed protocol, the home address request is directly made to the 
target host.

\item Publishing a FQDN on the Web: Alice can publish her FQDN on the
Web and distribute a different disposable home address to each 
correspondent that request a home address. This is advantageous over 
publishing a disposable home address directly, in which case multiple 
correspondents would obtain the same disposable home address. If that 
address is disposed of, all these correspondents would lose contact 
with Alice.

\end{enumerate}

An attacker who learned Bob's FQDN and attacking the his prime home 
address can be defeated by blocking that address address. In this case, 
the proposed protocol will be temporarily disabled. Correspondents 
that were previously distributed disposable home addresses can 
still reach Bob, however. Home address requests can still be made by 
verbal communication, e-mail, or instant messaging (clearly, in this
case, the users should be aware of disposable home addresses). In 
Section \ref{sec:sim}, we discuss this attack's efficiency in more 
details.

\subsection{Human interaction proofs}

When Alice requests a home address from Bob's device, a message 
will be displayed on Bob's device's screen, asking permission to return the
requested information. This request will interrupt the target
user and we see here a potential abuse. 

A malicious user who obtained a target user's FQDN may repeatedly make 
fake requests over the Internet in order to disturb the 
target user, forcing them to repeatedly pushing on NO button and 
eventually disabling the protocol. This attack would be similar to 
calling the target user and quickly hanging up. The attacker does not 
obtain anything, but disturbs the target user.  Consequently, in such 
cases, we would like to ensure that the initiator user makes some 
relatively harder work compared to the burden put on the target user, 
i.e. checking the request and making a decision. 

In order to prevent spam bots from obtaining a large number of e-mail 
accounts, HIP (Human Interaction Proof) tests also known as CAPTCHA 
(Completely Automated Public Turing test to tell Computers and Humans Apart)
are used today\cite{CAPTCHA}. As malicious character recognition programs become 
increasingly intelligent, the HIPs also become more and more 
difficult to read by humans. There is anecdotal evidence that some 
users do not comment on blogs when they are required to solve a 
HIP. Therefore, in our case, HIPs are good candidates to 
prove hard human work and defeat attackers who try to disturb users 
with bogus requests displayed on their screen. In the proposed 
protocol, when frequent home address requests are received by the 
target device, a HIP is returned to the initiator user, and before a 
notification is displayed on the target screen, the initiator 
must first present the correct answer. 

The difficulty of the HIP problem can be adaptively increased if
the requests are too frequent, which is a sign of anomaly. For 
example, longer CAPTCHAs can be returned.

\subsection{Simulations}
\label{sec:sim}

In this section, we present a simulation analysis of the above described 
home address distribution protocol. In the presented scenario, Alice, 
upon buying a new mobile phone, publishes her FQDN on the Web and has 
no contact initially. She has 200 potential correspondent users
which find Alice's FQDN on the Web, make a disposable phone number request
from Alice and make a call. Unfortunately, however, an attacker also learns 
Alice's FQDN, and hence her prime home address, and mounts battery exhaustion 
DoS attacks against Alice. The attack consists of continuously flooding 
the target phone with bogus requests during several hours. We assume that 
when under attack the prime home address is disabled and consequently 
correspondents' disposable home address requests are rejected. Consequently, 
the attack's impact is not battery consumption, but rejecting incoming 
calls from legitimate correspondents. Correspondents who obtain a disposable 
phone number (when there is no attack) can make further calls even if 
the prime home address is disabled. 

In the first simulation, the attacker mounts a 4-hour attack against 
Alice's mobile phone, everyday, where the attack begins at 8:00, 12:00 or 
16:00. During a day, the probability of making a call to Alice 
is \(\frac{1}{200}\) for each of 200 correspondents and 
the calls are made between 8:00 and 20:00, i.e. during one of the above
three attacks periods. Thus, the probability that a call coincides with an attack 
is \(\frac{1}{3} \times \frac{1}{3}= \frac{1}{9}\). If a disposable phone
number was not previously obtained and the call coincides with an attack
(i.e. the prime home address is blocked), the call is rejected. In the
second simulation, the attacker mounts a 6-hour attack which begins
at 8:00 or 14:00. In this case, the probability of rejecting a call is
\(\frac{1}{2} \times \frac{1}{2}= \frac{1}{4}\).

\begin{figure}
	\centering
	\subfigure[4-hour attack each day]{\includegraphics[width=\linewidth]{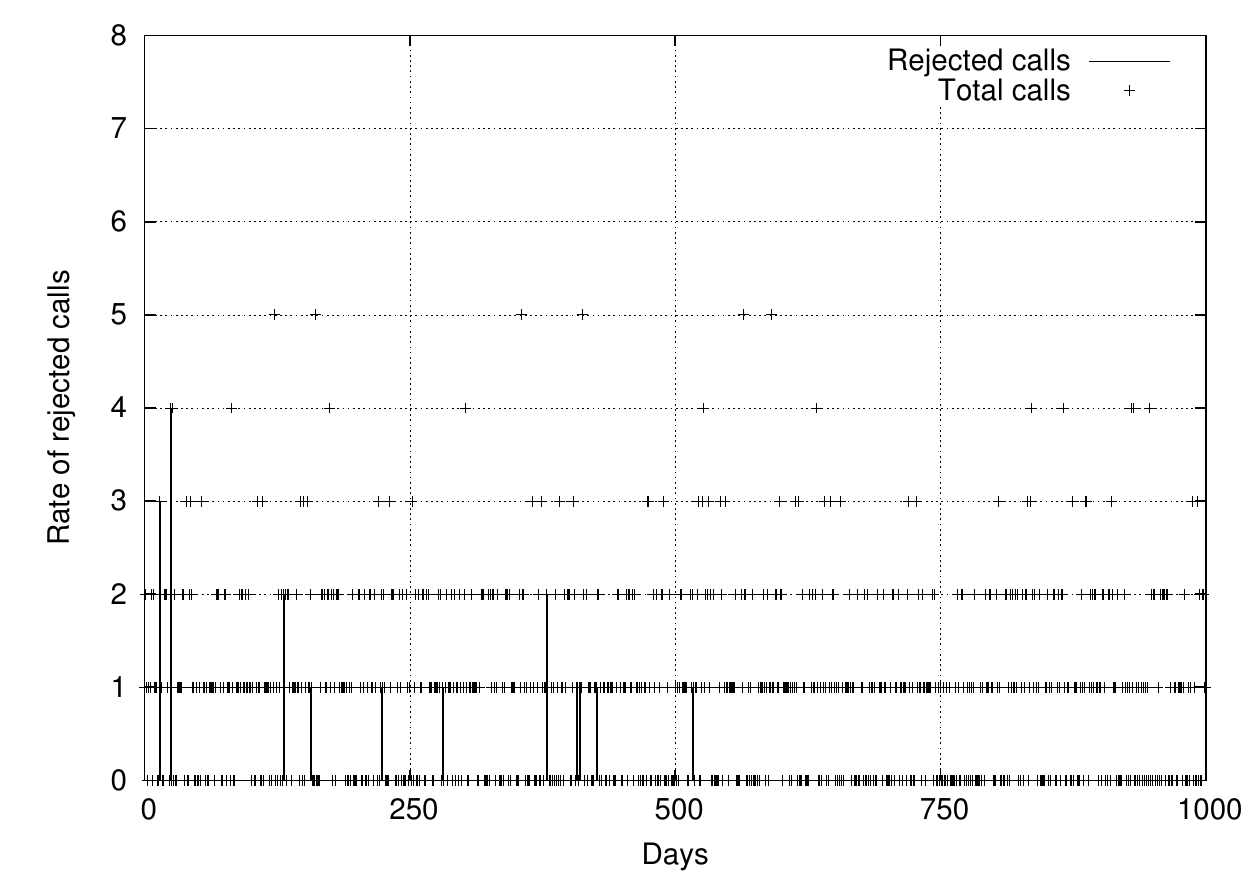}}
	\subfigure[6-hour attack each day]{\includegraphics[width=\linewidth]{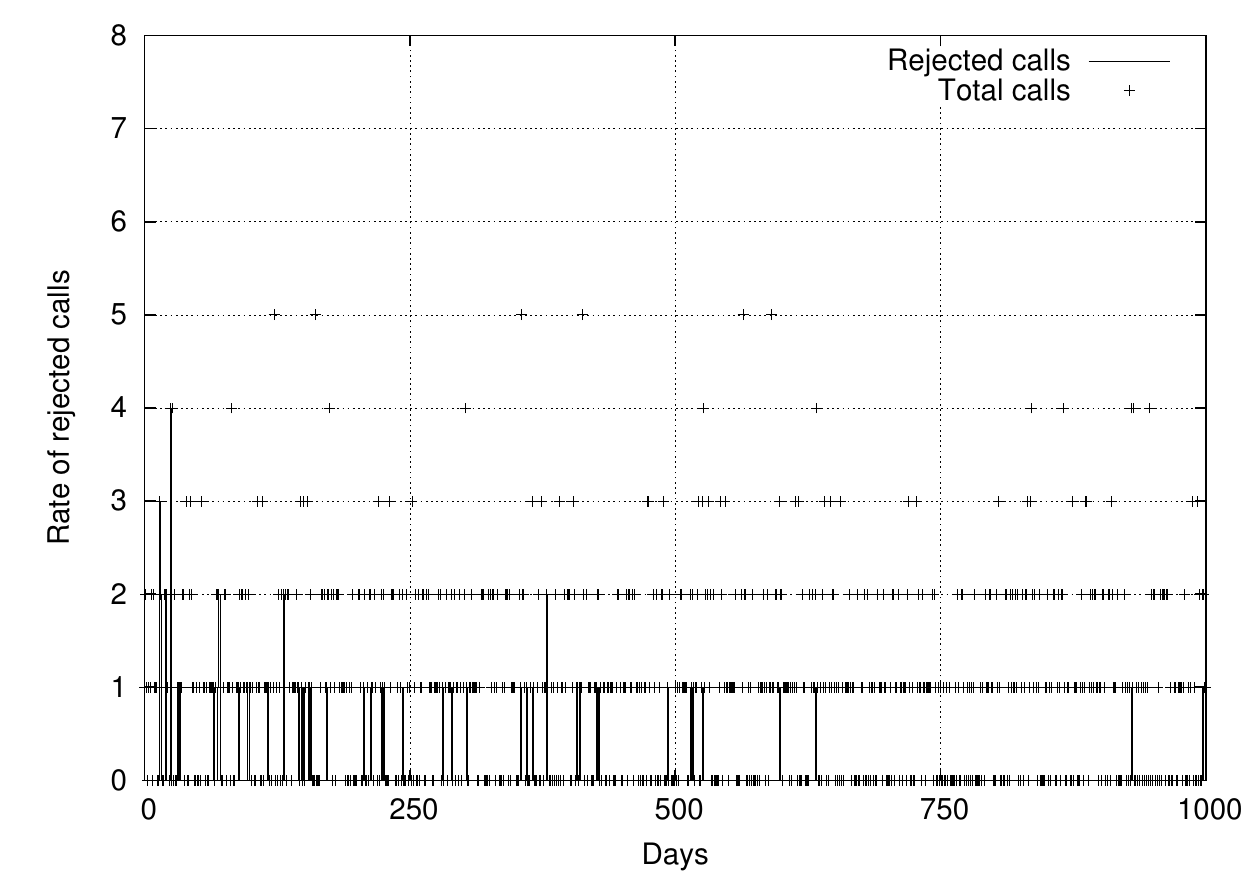}}
	\caption{Rate of rejected calls during 1,000 days of attack (hypothetical).}
	\label{fig:sim}
\end{figure}

The simulation results are shown in Figure \ref{fig:sim}. In the first 
simulation, a few calls are rejected, where some calls are still being 
rejected after 250 days. A total of 18 out of 1,993 calls are rejected
(0.9\%). In the second simulation, more calls are rejected and even 
after 500 days a few calls are being rejected. A total of 56 out of 
1,993 calls are rejected (2.8\%). 

We believe however that better results can be expected in practice. 
In practice, correspondents can also obtain disposable phone numbers 
through other channels e.g. user contact, e-mail, or instant messaging 
etc. The only impact of the attack would be therefore forcing a small 
portion of the correspondents (the rejected ones in Figure 
\ref{fig:sim}) to use e-mail, or instant messaging for requesting 
disposable home addresses. This is not as user friendly as using the
described home address distribution protocol but still acceptable.
Consequently, we believe that this attack has a low 
\textit{damage/effort ratio}\footnote{We define the damage/effort 
ratio as the measure of the seriousness of the damages caused by an attack
compared to the attacker's efforts. In this attack, negligible damage
is caused by mounths of persistent attack. Hence, it has a low 
damage/effort ratio.} and hence is probably unrealistic in practice. 

\section{Certificateless operation}
\label{sec:certificateless}

So far we have assumed that a Public Key Infrastructure (PKI) is 
available. When Alice makes a disposable home address request to Bob, 
the request is signed with Alice's private key, and Bob's answer is 
also signed with Bob's private key and certificates are attached to the
messages. Consequently, Bob is certain that the request was made by 
Alice, and Alice is certain that the returned home address belongs to 
Bob. 

However, if a PKI is not available, the attached certificates will not
be verifiable and consequently an attacker can impersonate Alice 
and/or Bob. For example, by impersonating Alice, an attacker can obtain
a home address and attack Bob. Or, by impersonating Bob, and attacker
can return her own address and act as a man-in-the-middle.

In this section, we propose a solution to address the case where a PKI 
is not available.

\subsection{User contact available}
\label{sec:contactok}
\begin{figure}[t]
	\centering
 	\includegraphics[width=7cm]{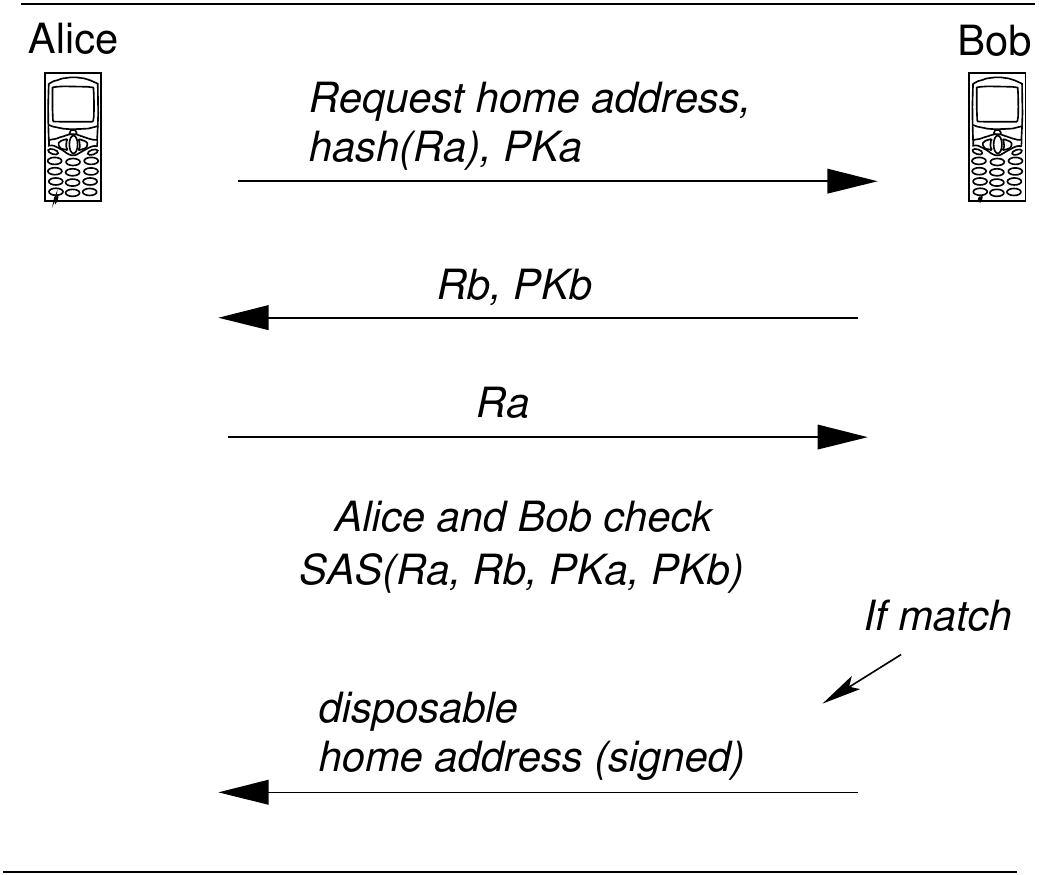}
	\caption{Protocol description (certificates not available)}
	\label{fig:protocol2}
\end{figure}

When user contact is available, i.e. both users are present, the 
protocol that is illustrated in Figure \ref{fig:protocol2} can be used 
for exchanging disposable home addresses and authentic certificates 
for future use. 

Here, we use the theoretical mutual authentication protocol described in 
\cite{SAS}. When Alice makes a home address request to Bob (by typing his 
FQDN), her device first generates a long random string \(Ra\), computes a 
secure hash \(hash(Ra)\) and sends it along with her public key \(PKa\). 
Bob's device generates its own random string \(Rb\) and sends it along 
with Bob's public key \(PKb\). Alice's device responds by sending 
\(Ra\). Bob's device computes the secure hash of \(Ra\) and compares 
it with the previously received \(hash(Ra)\). If they do not match, 
Bob's device aborts (not shown). Otherwise it continues. Alice's 
device computes and displays a SAS (short authentication string)
(\(n\)-bit hash, where \(n=15...20\)) out of \(Ra\),
\(Rb\), \(PKa\) and \(PKb\). Bob's device computes and displays the 
same. Alice reads the SAS to Bob through verbal communication (since 
both users are present, i.e. user contact is available). Alice and Bob 
compare the received SAS with the one their device computed and 
displayed. If they do not match, they abort (not shown). Otherwise 
the devices store each other's public keys, and securely exchange home 
addresses. 

This protocol is secure against man-in-the-middle attacks. 
Refer to \cite{SAS} for a security proof of the protocol\footnote{This 
protocol is actually a solution to the public key
distribution problem without PKI. Users, however, are generally not
aware of public keys. By integrating this solution into a phone number
(disposable home address in our case) exchange protocol, we also 
contribute to better security in mobile IP telephony. Users are more 
likely to run a contact management protocol than the pairing protocol 
described in \cite{SAS} alone. In this approach, phones get paired
when phone numbers are exchanged.}.

Since Alice and Bob securely exchanged their public keys, they can 
update disposable home addresses later when there is no longer user 
contact. For example, when an attacker learns the disposable home 
address that was given to Alice, Bob can block that address and Alice 
can request a new one.

\subsection{User contact not available}

If user contact is not available, certificateless operation is not 
secure in the absence of a PKI as described above. In this case, the 
users can use e-mail or instant messaging to exchange home addresses 
since these applications are password protected.

\section{Fighting SPIT}
\label{sec:spit}

The proposed model of operation can also be used against SPIT (SPam over IP Telephony). 
If a user that was given the disposable home address \(hoa_i\) 
uses it for sending unwanted messages or make disturbing phone calls, 
the address \(hoa_i\) can be blocked. The malicious user may request 
another address, however the target user can reject the request. 

For example, this model can be used to defeat disturbing calls and 
messages from a known user e.g. someone met in a party or 
telemarketing calls from a previously met salesman who was given a 
disposable home address.

\section{Architectural considerations}
\label{sec:arch}
In the proposed model of operation, a mobile host (and its user)
is identified using a FQDN, which can be viewed as a phone number.
The FQDN is resolved to the host's prime home address and disposable 
home addresses are requested from that address. A host can change its 
prime home address address at any time and register it to the DNS 
using dynamic DNS\cite{DYNDNS}. Disposable home addresses can 
also be viewed as phone numbers (disposable ones) since a user can use 
them directly. 

It can be noted that SIP URIs in the form of user@provider are not 
used\cite{SIP}. End-to-end SIP can be used for session establishment
(using IP addresses or FQDNs as SIP URIs), however the proposed model 
is independent from the SIP architecture, i.e. the SIP proxies are not 
needed. Mobility is managed by MIPv6. Since it does not rely on SIP 
infrastructure, the proposed model is more reliable against single 
points of failure.

Our proposal is however backward compatible, i.e. some users may use
the standard model using SIP infrastructure using a fixed single home
address at the risk of battery/CPU exhaustion.

One problem with our proposal is offline messaging. In cellular systems,
when a target mobile terminal is not online, the caller can leave
a message. The message is stored by the network and delivered when the
target becomes online again. For this, application-layer solutions 
are generally necessary, however in our case we do not use application-layer
infrastructure. No infrastructure is used other than the Mobile IPv6 
home agent and the DNS, and an offline messaging system cannot be 
possibly integrated into these systems. One solution to this problem 
is to deploy a peer-to-peer SIP telephony protocol as described in 
\cite{P2PSIP}. In this approach, the initiator records a voice message
and uploads copies of its message to a number of hosts that are online. 
Each host can then periodically try to forward the message to the 
target host until it becomes online. Details, e.g. optimizations, of 
this solution fall out of this paper's scope.

\section{Limitations}
\label{sec:lim}

In this paper, we are primarily concerned with remote attackers 
located anywhere in the Internet, attacking a particular known victim's
globally reachable home address. This, we believe, is the most serious
case since attacks are possible regardless of the victim's location.

However, an attacker located in the same cell as the victim, can also 
learn its MAC address and consume its energy by continuously sending 
packets to that address. Our solution cannot cope with these attacks. 
However, upon moving to a new cell, the victim can escape from the 
attacker.

The care-of-address may reveal the identity of a mobile user if the 
interface ID part of the address is constructed from the globally
unique MAC address. \cite{IPv6PRIV} describes privacy extensions for
IPv6 where the interface identifier is randomly generated. An attacker
may however decide to attack random care-of-addresses in the same 
subnet. Our solution cannot cope with these attacks. Similarly, the 
victim can escape from the attacker upon moving to a new subnet or by
configuring a new care-of-address.

\section{Implementation considerations}
\label{sec:imp}
Our problem in this paper is basically a home address management 
problem. It can be implemented at the application layer. The home address
management protocol that we describe will run between the mobile 
host's contact manager and a server application run by the home agent.
In this approach, the user makes a home address generation request to 
its contact manager which sends the request to the home agent. The 
home agent generates an address and configures it for the mobile host. 
Detecting address collisions is not a problem since the home agent 
knows all home addresses in its network. The trafic between mobile 
host and home agent is protected using IPsec, since they have a security
association.

The contact manager application on the mobile host should periodically 
monitor the mobile host's trafic, and run intrusion detection 
algorithms\footnote{In this paper, we do not enter into the details of
host-based network intrusion detection algorithms that can be used
by the mobile host, since this is a much broader topic which falls out 
of this paper's scope. Different implementations may use different
techniques.}. Upon suspicious activity, it should warn the user 
and upon user command, block the subject home address by sending a 
request to the home agent. The home agent should deconfigure the home 
address. Packets sent to that home address should no longer be served 
and they should be silently dropped by the home agent. The host should 
also deconfigure the home address.

Upon blocking the home address, the mobile host needs to configure
a new care-of-address and send to its home agent, otherwise the 
attacker can still reach the mobile host at its care-of-address (if
route optimization is used). If bidirectional tunneling method is
used, changing the care-of-address is not necessary since the 
care-of-address is hidden from the attacker.

In this approach, the contact manager applications on different devices 
also communicate with each other in order to request disposable home 
addresses using the protocols described in Section \ref{sec:dist}.

\section{Related work}
\label{sec:related}

\subsection{Disposable identifiers}

Our proposal was inspired by disposable phone numbers that are in use 
today\cite{DISPHONE}\cite{TOSSABLE}. 
Disposable phone numbers are used for coping with unwanted
correspondent users. Each user is given a different disposable phone number
and if one of the users show malicious behavior (e.g. make unwanted calls)
the phone number that was given to that user is canceled. 

Disposable e-mail addresses are also in use today, as an alternative 
way of sharing and managing e-mail addressing\cite{DISEMAIL}. 
A user distibutes a 
different e-mail address to each correspondent user and if anyone 
compromises it for any e-mail abuse, the address-owner can easily cancel it 
without affecting any other contact. 

In our case, i.e. in IP telephony, lower level threats also exist, e.g.
CPU/energy exhausting DoS attacks as discussed throughout the paper. Consequently,
we applied this disposable identifier solution at the IP layer, by proposing
disposable IP addresses, i.e. disposable home adresses. This way we can 
cope with both problems, CPU/energy exhaustion DoS attacks and SPIT, using
the same solution.

\subsection{Battery/CPU exhaustion attacks}

In \cite{DUCKLING}, the authors warn about so called ``sleep deprivation'' attacks
in the context of mobile ad hoc and sensor networks. This attack consists 
of periodically sending packets to a victim in order to prevent it from 
sleep mode. 

In \cite{BATTERYDOS}, the authors offer a thorough analysis of denial-of-service
attacks on battery-powered mobile computers. In this paper, three main 
attack methods are described: (1) service request attacks, where repeated
requests are made to the victim for services, typically over a network,
(2) benign power attacks where the victim is forced to execute a valid but 
energy hungry task repeatedly, and (3) malignant power attacks where the
attacker modifies or creates an executable to make the system consume
more energy than it would otherwise. Our work is different in that
we assume that the attacker cannot obtain access to the victim host, 
and services like ssh are deactivated. In our case, the victim consumes 
energy for receiving request packets and returning replies. A session
is never established. Energy is consumed although the attacker does not 
obtain access to the target mobile phone.

Key agreement protocols e.g. IKEv2 which employ the Diffie-Hellman
protocol, are vulnerable to CPU exhaustion attacks since they rely
on CPU expensive operations like exponentiation. An attacker can 
repeatedly make key agreement requests forcing the victim to 
continuously make CPU expensive operations. A well-known solution
to these attacks is \textit{client puzzles}\cite{PUZZLES}. Upon receipt of the 
initiator's key agreement request, the responder returns a puzzle. 
The puzzle is difficult to solve but the proposed result is easy to 
check. For example, the responder returns a nonce \(N\) and the 
initiator is requested to find a solution \(X\) such that the first
\(k\) bits of \(hash(X | N)\) are zero. The only way to find the 
solution is brute force, i.e. trying all possible solutions. The
solution is easy to check and consists of computing \(hash(X, N)\).
Consequently, the attacker makes much more work than the victim, which 
slows down the attack. This solution can be used to defend against CPU 
exhaustion attacks, but energy is still consumed for receiving packets 
(deprived from power save mode) and responding with puzzle requests. 
In addition, an attacker may orchestrate a DDoS attack using tens of
thousands of compromized hosts and consequently use much higher CPU 
power than the victim(s). I.e., each host participating in the attack 
may need longtime to solve the requested puzzles, but the overall 
attack performance may be good enough to exhaust the CPU/battery of 
victims(s). A different solution e.g. disposable home addresses is 
therefore necessary. 

\subsection{SPIT}
\label{sec: relatedSPIT}
\cite{SIPSPAM} provides a detailed analysis of the SPIT threat in the context
of SIP. The authors identify three threats: Call spam, instant messaging
spam and presence spam. They also analyze the solution space and 
propose various solutions e.g. content filtering, black lists, white
lists, reputation systems, limited use addresses etc. Among these solutions
limited use addresses are very similar to our proposal. The authors do not 
propose a solution for SIP, but instead they bring to our attention 
the use of limited use addresses for protection against e-mail spam, where
each correspondent user is given a different e-mail address. 

Our proposal is different in that we use disposable IP addresses in 
order to overcome not only the SPIT problem but also the battery/CPU 
exhaustion threats. We agree however that more than one solutions can 
be applied to the SPIT problem.

\subsection{Pseudo home addresses}

In \cite{LOCPRIV} the authors describe the use of a \textit{pseudo 
home address} to achieve location privacy. A pseudo home address is 
used to replace the real home address in various messages, which 
allows the mobile node to hide its real home address from both the 
correspondent node and eavesdroppers. 

\subsection{Network ingress filtering}

An alternative solution against Internet attacks is \textit{network 
ingress filtering}, which consists of dropping packets which have
topologically incorrect source address\cite{INGRESS}. We believe that 
this is a good solution in theory, however note that most service 
providers to not employ ingress filtering. This solution is also not 
effective against DDoS attacks in which the attacker controls a large 
number attacking hosts which send packets with their real IP address.

\section{Conclusion}
\label{sec:conc}

In this paper we proposed disposable home addresses for Mobile IPv6 
(MIPv6). In summary, the proposed solution will help defend against 
the following security problems in future mobile IP telephony:

\begin{itemize}
\item Battery exhaustion Denial-of-Service (DoS)
attacks that deprive a target host from sleep
mode (important since mobile hosts are battery powered and energy is
a scarce resource)

\item Battery exhaustion DoS attacks that force a target host to return
reply packets and link-layer ACKs

\item CPU exhaustion DoS attacks that force a target to process incoming 
malicious packets and return replies

\item CPU exhaustion DoS attacks that force the target to perform 
CPU-expensive cryptographic operations in IKEv2 (unlike client 
puzzles, defense is possible against DDoS attacks)

\item Disturbing calls and messages from a known user e.g. someone 
met in party or telemarketing calls from a previously met salesman 
(subset of the SPam over IP telephony problem)

\item Other attacks e.g. port scanning, or currently unknown attacks 
may also be defeated

\end{itemize}

These defenses are achieved by distributing a different ``disposable''
home address to each correspondent. When one of the addresses
is exploited for an attack, it is disposed of by its owner i.e. the
mobile user. Other correspondents that were distributed different
home addresses can still reach the user. It is also possible to 
request a new disposable home address when one of them is disabled.

It is noteworthy that the proposed model of operation is only
possible in IPv6 since a large number of mobile hosts may be served
by the home network and each mobile host may have hundreds of disposable
home addresses. In IPv6, \(2^{64}\) addresses are supported in a same
subnet, which is much more than enough.

The proposed model is especially useful against battery exhausting 
DoS attacks and CPU exhausting distributed DoS attacks, since it seems 
to be the only viable solution, currently. We believe that these attacks
are fundamentally difficult to defeat and represent a considerable 
threat against comfortable use of future wireless Internet. Without
the solution proposed in this paper, the only defense seems to be turning 
off the phone that is under attack, for later use when the attack ceases. 

\bibliographystyle{IEEEtran}
\bibliography{bibliography}

\end{document}